\documentclass[aps,prb,reprint,superscriptaddress,amsmath]{revtex4-1}
\usepackage{hyperref}
\usepackage[dvips]{graphicx}
\usepackage{amsmath,amssymb,amsbsy}

\newcommand{\txt}[1]{\mathrm{#1}}

\newcommand{\EC}{E_\txt{c}}

\newcommand{\kB}{k_\txt{B}}

\newcommand{\unit}[1]{\ \mathrm{#1}}

\newcommand{\Vds}{V_\mathrm{ds}}

\newcommand{\Idet}{I_\mathrm{det}}

\newcommand{\D}{\,\mathrm{d}}

\newcommand{\nS}{n_\mathrm{S}}
\newcommand{\nqp}{n_\mathrm{qp}}
\newcommand{\fS}{f_\mathrm{S}}
\newcommand{\fN}{f_\mathrm{N}}

\newcommand{\RT}{R_\mathrm{T}}

\renewcommand{\Re}{\mathrm{Re}}

\newcommand{\TS}{T_\mathrm{S}}
\newcommand{\TN}{T_\mathrm{N}}

\newcommand{\be}{\begin{equation}} \newcommand{\ee}{\end{equation}}
\newcommand{\ba}{\begin{eqnarray}} \newcommand{\ea}{\end{eqnarray}}

\newcommand{\ie}{i.\,e.}

\begin{document}

\title{Vanishing quasiparticle density in a hybrid Al/Cu/Al single-electron transistor}
\author{O.-P. Saira}
\affiliation{Low Temperature Laboratory, Aalto University, P.O. Box 15100, FI-00076 AALTO, Finland}
\author{A. Kemppinen}
\affiliation{Centre for Metrology and Accreditation (MIKES), P.O. Box 9, 02151 Espoo, Finland}
\author{V. F. Maisi}
\affiliation{Centre for Metrology and Accreditation (MIKES), P.O. Box 9, 02151 Espoo, Finland}
\author{J. P. Pekola}
\affiliation{Low Temperature Laboratory, Aalto University, P.O. Box 15100, FI-00076 AALTO, Finland}

\begin{abstract}
The achievable fidelity of many nanoelectronic devices based on superconducting aluminum is limited by either the density of residual nonequilibrium quasiparticles $\nqp$ or the density of quasiparticle states in the gap, characterized by Dynes parameter $\gamma$. We infer upper bounds $n_\mathrm{qp} < 0.033\unit{\mu m^{-3}}$ and $\gamma < 1.6\times10^{-7}$ from transport measurements performed on Al/Cu/Al single-electron transistors, improving previous results by an order of magnitude. Owing to efficient microwave shielding and quasiparticle relaxation, typical number of quasiparticles in the superconducting leads is zero.
\end{abstract}
\maketitle

\section{Introduction}

Active research and debate on the origin and density of residual quasiparticles in aluminum-based superconducting quantum circuits prevails currently~\cite{Catelani_relaxation, *Leander_decay, Martinis_noneqQP}. Aluminum is a widely used metal in the field of low-temperature mesoscale electronics due to its superconducting properties and its tendency to form a native oxide that can be employed as a tunnel barrier. The practical performance of superconducting devices is often degraded by excess quasiparticle processes that do not follow from the assumption of full thermal equilibrium and Bardeen-Cooper-Schrieffer (BCS) form for the quasiparticle density of states of the superconducting electrodes. The figures of merit that we will address in this work are the density of non-equilibrium quasiparticles $\nqp$ and the Dynes parameter $\gamma$ for the normalized density of quasiparticle states in the gap. These parameters are presently major limiting factors for the coherence time of Josephson junction qubits~\cite{Martinis_noneqQP, Shaw_Kinetics, Makhlin_Engineering}, relaxation time of highly sensitive radiation detectors~\cite{deVisser_qpfluct, *Barends_2008, *Barends_2009}, the ultimate temperature reachable by normal-insulator-superconductor (NIS) junction refrigerators~\cite{Giazotto_RMP}, and potentially for the accuracy of the SINIS turnstile~\cite{Pekola_subgap}, a contender for the realization of a metrologically accurate source of quantized electric current. 

The lowest values for $\gamma$ that have been obtained from subgap conductance measurements of NIS junctions and SINIS single-electron transistors (SETs) are $2 \times 10^{-5}$ and $1 \times 10^{-6}$, respectively~\cite{Pekola_subgap}.  Subgap conductance of defect-free NIS junctions is dominated by two-particle Andreev tunneling~\cite{Greibe_Pinholes2011}. For $\nqp$, studies on superconducting qubits \cite{Martinis_noneqQP, Shaw_Kinetics} and resonators \cite{deVisser_qpfluct} have reported a low-temperature saturation to $10-55 \unit{\mu m^{-3}}$ due to an unidentified excitation mechanism. Recently, a significantly improved result $\nqp < 1\unit{\mu m^{-3}}$ has been obtained by a transmon qubit realization~\cite{Paik_coherence}. A similar number can be inferred from the quasiparticle tunneling rates reported in Ref.~\onlinecite{Court_qptraps} for a superconducting single-electron transistor (SET) design with normal metal quasiparticle traps. In this work, we present quasiparticle transport measurements of Al/AlOx/Cu SETs combining carefully implemented  shielding against external black body radiation and quasiparticle traps. The results yield unprecedented upper bounds $\nqp < 0.033 \unit{\mu m^{-3}}$ and $\gamma < 1.6\times10^{-7}$.

\begin{figure}[tb!]
    \begin{center}
        \includegraphics[width=.49\textwidth]{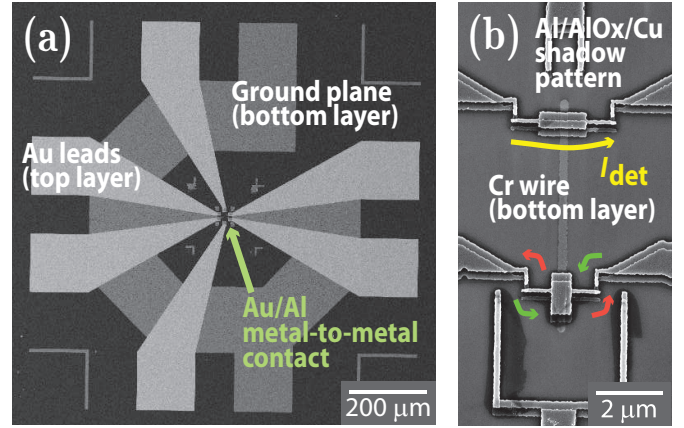}
    \end{center}
   \caption{\label{fig:1} (a) Large scale electron micrograph of a sample showing the electrical connections to the active region in the middle. Portions of the ground plane, covered by an insulating AlOx layer of 25~nm thickness, are also visible. (b) Active area of one of the measured samples, consisting of two Al/AlOx/Cu/AlOx/Al SETs that are coupled capacitively by a Cr wire located underneath the AlOx layer. Colored arrows illustrate the monitored in- and out-tunneling events (green and red, respectively), and the macroscopic electro\-meter current (yellow) in a configuration where the upper SET is used as the electrometer. Shadow-evaporated leads terminate at ohmic Au/Al contacts beginning $10\unit{\mu m}$ away from the junctions (not shown).}
\end{figure}

The experiment was performed on samples similar to that pictured in Fig.~\ref{fig:1}. We monitored the charge state on the central island of a hybrid, \ie, SINIS-type, SET using another hybrid SET as an electrometer. The electrometer current $\Idet$ was read out using a room-temperature current amplifier in the so-called DC-SET configuration. Most of the capacitive coupling between the SETs was provided through a $7\unit{\mu m}$ long Cr wire galvanically isolated from the SET metal layers.
Following the technique introduced in Ref.~\onlinecite{Pekola_subgap}, all electrical leads of the $1\times 1\unit{mm^2}$ pattern on the chip from the bonding pads to the active region were capacitively shunted by a conducting groundplane that was electrically isolated from the leads by the 25~nm AlOx layer. Pathways for quasiparticle trapping from the aluminum electrodes are provided by the overlap through the tunnel barrier oxide to the copper electrodes extending to within few hundreds of nm of the tunnel junctions, and the ohmic contact between Al and Au films beginning 10$\unit{\mu m}$ away from the junctions. We also fabricated and measured a reference sample without the ohmic Al/Au contact, but having an otherwise equivalent design. 

\section{Theoretical description of single-electron processes}

The basis of our theoretical modeling of the charge transfer is the Golden rule expression for the first-order tunneling rate
\begin{multline}
\Gamma^\txt{1e}(E) = \frac{1}{e^2\RT}\int_{-\infty}^\infty\D E_1 \int_{-\infty}^\infty\D E_2 \nS(E_1) \fS(E_1)\\
[1-\fN(E_2)] P(E_1-E_2+E), \label{eq:goldenrule}
\end{multline}
where $\nS(E) = \left| \Re \frac{E}{\sqrt{E^2 - \Delta^2}} \right|$ is the quasiparticle density of states in the superconducting electrode, $\fS$ ($\fN$) is the occupation factor in the superconducting (normal) electrode, and $P(E)$ is the probability to emit energy $E$ to the electromagnetic environment during the tunneling. The occupation factors are taken to be Fermi functions at temperature $\TN$ ($\TS$), \ie, $f_\txt{N,S}(E) = \left[1 + \exp(E/\kB T_\txt{N,S})\right]^{-1}$. The Golden rule formula with $P(E) = \delta(E)$, equivalent to a zero-impedance environment, has been succesfully used to describe a wide range of charge and energy transport phenomena in SINIS structures \cite{Saira_HeatTransistor} in the range $E \gtrsim \Delta$. Next, we will consider theoretically the mechanisms that can cause excess quasiparticle processes in the subgap range $E < \Delta$.


\emph{(i)} Even if charge transport is completely described by the above model, quasiparticle thermalization in both the N and S electrodes can be nontrivial at sub-kelvin temperatures due to strongly suppressed electron-phonon coupling. In a later section, we demonstrate that quasiparticle temperature $\TN$ does not deviate significantly from the bath temperature $T_0$ at temperatures above 50~mK. On the other hand, we characterize excess quasiparticle excitations in the superconductor by the density of nonequilibrium quasiparticles
\begin{equation}
\nqp = 2 D(E_F) \int_\Delta^{\infty} \D E \nS(E) \fS(E), \label{eq:nqp}
\end{equation}
where $D(E_F)$ denotes the density of states at the Fermi energy. We use the literature value\cite{Court_qptraps} $D(E_F) = $ \mbox{$1.45 \times 10^{47}\unit{m^{-3} J^{-1}}$}. At base temperature, we have $\Delta / (\kB \TN) \sim 50$, and hence the induced quasiparticle tunneling in the gap assumes a bias-independent rate
\begin{equation}
\Gamma^\txt{1e}_\txt{nqp} = \frac{\nqp}{2 e^2 \RT D(E_F)}. \label{eq:Gamma_nqp}
\end{equation}

\emph{(ii)} Experimentally observed finite subgap conductance in NIS junctions is ofted modeled by introducing the life-time broadened Dynes density of states~\cite{Dynes_qp_lifetime, *Dynes_tunneling} for the quasiparticle excitation spectrum of the S electrode
\begin{equation}
\nS(E) = \left|\Re \frac{E/\Delta + i \gamma}{\sqrt{(E/\Delta + i \gamma)^2-1}}\right|, \label{eq:Dynes}
\end{equation}
where parameter $\gamma$ effectively expresses the quasiparticle density of states in the middle of the gap as a fraction of the density in the normal state. Equation~(\ref{eq:Dynes}) may not be valid far from the gap edges~\cite{Mitrovic_DOS}, and we stress that in the present work the Dynes model is only used as a tool to assess the effect of subgap states on quasiparticle tunneling. 

\emph{(iii)} The $P(E)$ function appearing in Eq.~(\ref{eq:goldenrule}) can be calculated from the autocorrelation function of phase fluctuations over the junction \cite{IngoldNazarov}. In previous work \cite{Saira_EAT}, it has been explicitly demonstrated that photon assisted tunneling (PAT) due to microwave irradiation ($f \gtrsim \Delta/h = 50\unit{GHz}$) originating from outside the sample stage can be a dominant factor in the dynamics of metallic single-electron devices. In the present experiment, a known source of harmful phase fluctuations is the high-frequency component of detector back-action. It originates from the switching noise that results from loading and unloading the detector SET island as the probe current is transported through. In addition, blackbody radiation from higher-temperature stages of the cryostat can reach the junction due to insufficient filtering of the signal lines or leaks in the radiation shields enclosing the sample. The approximation $P(E) = \frac{\pi S_V(|E|/\hbar)}{R_\txt{K} E^2},$ valid for $E < 0$ and sufficiently weak $S_V$~\cite{Martinis_EA}, gives a straightforward relation between the absorptive part of the $P(E)$ and the power spectrum $S_V(\omega)$ of the voltage noise over the junction. The above-described detector back-action is an instance of random telegraph noise (RTN), for which the relevant high-frequency part of the power spectrum can be written as $S_V^\txt{det}(\omega) = \left(\frac{\xi(\omega) \kappa e}{2 C_\Sigma^\txt{det}} \right)^2 \frac{\left| \Idet \right|}{\pi e \omega^2},$ where $\kappa$ is the fraction of island charge coupled from the device under test (DUT) to the detector, $C_\Sigma^\txt{det}$ is the total capacitance of the detector SET island, and $\xi(\omega)$ describes high-frequency attenuation of the Cr wire~\footnote{The expression is valid for a detector with two active charge states assuming transition rates between them are much smaller than $\omega$.}.

\begin{figure}
    \begin{center}
        \includegraphics[width=.49\textwidth]{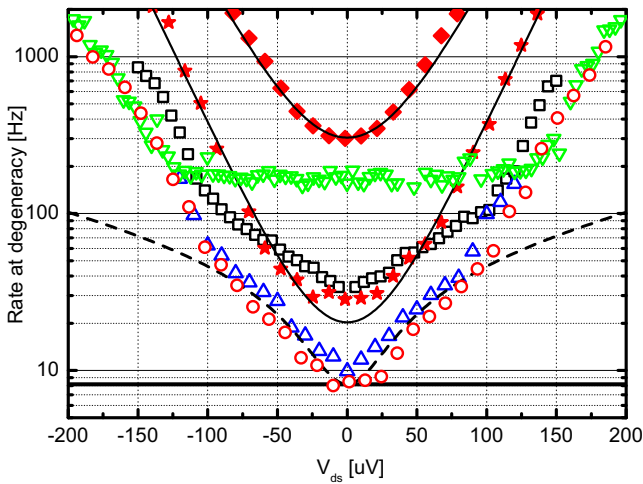}
    \end{center}
   \caption{\label{fig:2} Bandwidth-corrected single-electron tunneling rates as a function of $\Vds$ at charge degeneracy measured in different setups at base temperature: PT (18~mK, open upward triangles), PDR1 (50~mK, open circles), and PDR2 (50~mK, open squares). For PDR1, data from higher temperatures is also given: 131~mK (filled stars), and 158~mK (filled diamonds). All these measurements were performed with the same sample employing Au/Al contacts for quasiparticle trapping. The two thin solid lines represent thermally activated rates calculated for the known sample parameters at 158~mK and 135~mK. The dashed line is the theoretical rate for $\gamma = 1.6\times10^{-7}$ at $\TS$ = $\TN$ = $50~\unit{mK}$, and the horizontal thick line represents the rate induced by $\nqp = 0.033\unit{\mu m^{-3}}$. Open downward triangles: Base temperature data from a reference sample without Al/Au contacts measured in PDR1. For ease of comparison, tunneling rates from the reference sample have been scaled by the ratio of junction conductances $\frac{G_L + G_R}{G^\txt{ref}_L + G^\txt{ref}_R}$.} 
\end{figure} 

\emph{(iv)} Certain higher-order processes could appear in the experimental detector traces as single-electron processes. The processes in question are Andreev tunneling of a Cooper pair into the island followed by a rapid relaxation of a single quasiparticle through one-electron tunneling, and Cooper pair--electron cotunneling, which changes the charge on the island by one electron. Theoretical predictions for the rates can be made based on the results of Ref.~\onlinecite{Averin_PumpTheory} using the known sample parameters and a value $g/\mathcal{N} = 10^{-5}$ for the normalized conductance per channel \footnote{The estimate for $g/\mathcal{N}$ is extrapolated from values of Ref.~\onlinecite{Maisi_Andreev} by scaling with junction area and resistance.}. Predicted higher-order rates are orders of magnitude smaller than the experimental observations for bias voltages $|\Vds| < 100\unit{\mu V}$.

\begin{figure}
    \begin{center}
        \includegraphics[width=.49\textwidth]{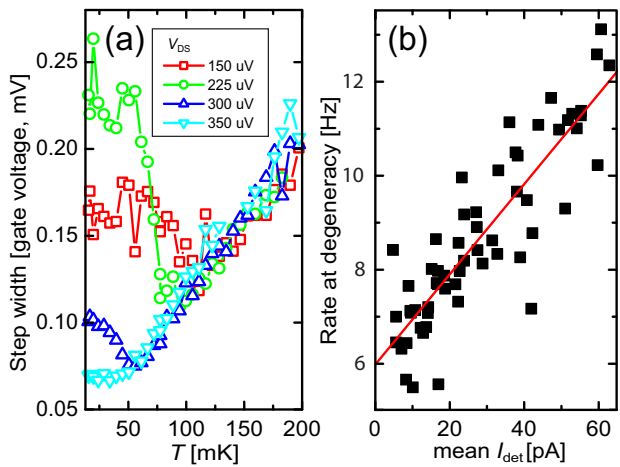}
    \end{center}
   \caption{\label{fig:3} (a) Measured width of the Coulomb staircase steps as a function of bath temperature for different bias voltages of the DUT. (b) Observed zero-bias tunneling rates at base temperature for different detector currents, and a linear fit to the data. The plotted current is the mean value of the detector trace from which the transition rate was determined.}
\end{figure} 

\section{Experimental methods}

To determine experimentally the single-electron tunneling rate for a fixed bias voltage $\Vds$, the gate voltages were adjusted so that the detector was at a charge sensitive operating point, while the DUT was at charge degeneracy. At degeneracy, owing to sufficient $\EC$ of the DUT, only the two degenerate charge states have a non-vanishing population. Hence, the expected electrometer signal is RTN, and the transition rate is given by $\Gamma = \Gamma^{1e}(eV_\txt{ds}/2) + \Gamma^{1e}(-eV_\txt{ds}/2)$, where two identical junctions have been assumed. The state transitions were identified from the recorded electrometer traces by digital low-pass filtering followed by a threshold detector. We used the analytical model of Ref.~\onlinecite{Naaman_Rates} to compensate for missed transitions due to finite detector bandwidth.

To test the coupling of microwave radiation to the junctions, we repeated the experiments in several cryostats equipped with different wiring and  shielding solutions. Two independent setups, henceforth denoted by PDR1 and PT, to yield the lowest tunneling rates, suggesting that the external microwave  radiation was suppressed to a negligible level. In both setups, standard solutions were used for the microwave filtering of the signal lines: 1~m of Thermocoax in PDR1 and a combination of Thermocoax and powder filters in PT. However, similar to other recent works~\cite{Barends_2011, *Kemppinen_2011}, the sample chip was protected against radiation shining from the higher-temperature parts of the cryostats by two nested rf-tight shields. The microwave shields were attached to the sample stage body by either threads sealed  with indium (PDR1), or by screws (PT). For illustration, we present data also from setup PDR2 that is similar to PDR1, but in which the Thermocoax lines terminate to a connector that was not sufficiently rf-tight.

\section{Results and conclusions}

Our main experimental data, the measured transition rates as a function of the device bias $\Vds$, are presented in Fig.~\ref{fig:2}. A saturation temperature below which the observed rates did not decrease was found around 80~mK. Using the value $\RT = 1.1\unit{M\Omega}$ ($2.0\unit{M\Omega}$ and $25.0\unit{M\Omega}$ for the junctions of the more asymmetric reference sample) obtained from device $I$-$V$ characteristic measured with an ordinary ammeter, we determined the value $\Delta = 210\unit{\mu V}$ by fitting the 158~mK data to the thermally activated (TA) rates assuming $\TN = \TS = T_0$. The large scale $I$-$V$ characteristic is consistent with this value. Data obtained at 131~mK agrees with TA predictions for $135\unit{mK}$ except in the range $|\Vds| < 25\unit{\mu V}$ where the TA rates become comparable to the saturation floor. In contrast, the theoretical TA rates for 50~mK are below 0.01~Hz in the $\Vds$ range that was accessible in the counting experiment. 

In order to aid the analysis of the base temperature results, we present two auxiliary results ruling out plausible explanations for the observed low-temperature saturation. First, to demonstrate the thermalization of the normal metal island to 50~mK, we study the width of the transition between two charge states as the gate charge of the DUT is swept past a degeneracy point, following the procedure of Ref.~\onlinecite{Saira_EAT}. The obtained step widths in PT cryostat are presented in Fig.~\ref{fig:3}(a).
For bias voltages far below the gap edge, anomalous broadening of the step at low temperatures is a clear indication of non-thermal processes. However, at the highest studied bias voltage $\Vds = 350\unit{\mu V}$, the linear TA regime extends down to 50~mK. Biasing of the DUT does not affect the temperature of the normal metal island as the electronic cooling power for $|\Vds| \leq 350\unit{\mu V}$ is negligible~\cite{Giazotto_RMP}.

Secondly, we studied experimentally the influence of detector back-action by measuring zero-bias rates at the base temperature at different operating points of the detector. A linear dependence of $\Gamma$ on $\Idet$ can be observed in the data from PDR1 cryostat presented in Fig.~\ref{fig:3}(b) as predicted by the $P(E)$ theory for the small detector currents employed here ($\Idet \ll e \Delta/h$). By extrapolating to zero detector current, we deduce that for the typical detector current of 30~pA in the time traces on which Fig.~\ref{fig:2} is based, the contribution from detector back-action is 2--3~Hz, which is a significant but not a dominating fraction.

The tunneling rate data from the reference sample without Al/Au contacts shown in Fig.~\ref{fig:2} displays a clear plateau consistent with a quasiparticle density of $\nqp = 0.69\unit{\mu m^{-3}}$. For this sample, the dominant quasiparticle relaxation channel at base temperature is tunneling through the oxide barrier to the normal metal shadow. This allows us to infer a homogenous injection rate of $r_\txt{qp} = 3\times 10^{5}\unit{s^{-1}\, \mu m^{-3}}$, presumably due to external radiation. Applying the quasiparticle diffusion model of Ref.~\onlinecite{Martinis_noneqQP} with a thermal energy distribution, the Al/Au contacts are expected to bring about a 100-fold reduction in $\nqp$ at the junctions. From the observed rates in PDR1 and PT setups, we can infer the previously stated upper bound $\nqp < 0.033\unit{\mu m^{-3}}$ for the sample with Al/Au contacts as illustrated in the figure, \ie, a 20-fold reduction. Given the nominal thickness of 30~nm for the aluminum film, we find that the expected number of quasiparticles in the total volume of one of the aluminum leads is less than 0.1, \ie, the superconductors are nearly free of quasiparticles in a time-averaged sense. 

However, based on the absence of a flat plateau in experimental bias dependence, non-equilibrium quasiparticles cannot account for a majority of the observed base temperature tunneling events. The curve for Dynes model with $\gamma = 1.6\times10^{-7}$ is a better match to the experimental results in the range $|\Vds| < 100\unit{\mu V}$, more so for the data from PT cryostat, but the present amount of data does not allow one to make a definite statement about the presence of subgap states at $\gamma \sim 10^{-7}$ level.

Finally, it should be noted that the base temperature tunneling rate at zero bias is four times higher in the imperfectly shielded PDR2 setup compared to those achieved in PDR1 and PT setups. Elevated rates in PDR2 are caused by PAT due to stray blackbody radiation. The results from PDR1 and PT are very similar, and part of the discrepancy between them can be attributed to uncertainty in the detector bandwidth compensation (in the notation of Ref.~\onlinecite{Naaman_Rates}, $\Gamma_\txt{det}$ = 2000~Hz for PDR1 and 125~Hz for PT) and different probing current. Also, for the high base temperature tunneling rates in bias range $100-200\unit{\mu V}$, we find PAT to be a plausible explanation. The observed rates can be reproduced by assuming a microwave noise spectrum with an exponential high-frequency cutoff corresponding to an effective temperature of $350\unit{mK}$, and a spectral density of the order of $1\unit{pV/\sqrt{Hz}}$ at $f = 50\unit{GHz}$.

In conclusion, we have demonstrated that microwave shielding and enhanced quasiparticle relaxation play a key role in achieving the highest possible performance of superconducting aluminum based devices. The possibility to employ these techniques in future realizations of quantum information processing devices presents exciting prospects.

\section*{Acknowledgements}

We acknowledge M.~M\"ott\"onen, C.~Flindt, P.~Delsing, and T. Aref for valuable discussions during drafting of the manuscript. This work has been supported by the Academy of Finland, V\"ais\"al\"a Foundation, the Finnish National Graduate School in Nanoscience, Technology Industries of Finland Centennial
Foundation, and the European Community's Seventh
Framework Programme under Grant Agreements No.~217257 (EURAMET joint research project REUNIAM),
No.~218783 (SCOPE), and No.~238345 (GEOMDISS).

\bibliography{electron_pump}

\end{document}